\documentclass[aps,pra,superscriptaddress,reprint,longbibliography]{revtex4-2}
\usepackage{graphicx}
\usepackage{amsmath}
\usepackage{amsfonts}
\usepackage{enumerate}
\usepackage{multirow}
\usepackage{makecell}
\usepackage[table,xcdraw]{xcolor}
\usepackage{times}
\usepackage[hidelinks]{hyperref}
\usepackage{soul}
\hypersetup{
    colorlinks,
    linkcolor={blue!80!black},
    citecolor={blue!80!black},
    urlcolor={blue!80!black}}
\usepackage[utf8x]{inputenc}

\begin{document}

\title{Synchronization and Non-Markovianity in open quantum systems}

\author{G\"oktu\u{g} Karpat}
\affiliation{Faculty of Arts and Sciences, Department of Physics, \.{I}zmir University of Economics, \.{I}zmir, 35330, Turkey}

\author{\.{I}skender Yal\c{c}\i nkaya}
\affiliation{Department of Physics, Faculty of Nuclear Sciences and Physical Engineering, Czech Technical University in Prague, B\v{r}ehov\'a 7, 115 19 Praha 1-Star\'e M\v{e}sto, Czech Republic}

\author{Bar\i\c{s} \c{C}akmak}
\affiliation{College of Engineering and Natural Sciences, Bah\c{c}e\c{s}ehir University, Be\c{s}ikta\c{s}, \.{I}stanbul 34353, Turkey}

\author{Gian Luca Giorgi}
\affiliation{IFISC (UIB-CSIC), Instituto de Fisica Interdisciplinar y Sistemas Complejos - Palma de Mallorca, E-07122. Spain}

\author{Roberta Zambrini}
\affiliation{IFISC (UIB-CSIC), Instituto de Fisica Interdisciplinar y Sistemas Complejos - Palma de Mallorca, E-07122. Spain}

\date{\today}

\begin{abstract}

Detuned systems can spontaneously achieve a synchronous dynamics and display robust quantum correlations in different local and global dissipation regimes. Beyond the Markovian limit, information backflow from the environment becomes a crucial mechanism whose interplay with spontaneous synchronization is unknown. Considering  a  model of two coupled qubits, one of which interacts with a dissipative environment, we show that non-Markovianity is highly detrimental for the emergence of synchronization, for the latter can be delayed and hindered  because of the presence of information backflow. The results are obtained considering both a master equation approach and a collision model based on repeated interactions, which represents a very versatile tool to tailor the desired kind of environment.

\end{abstract}

\maketitle

\section{Introduction}

Synchronization between different units, due to their interaction, is a paradigmatic phenomenon quite widespread in nature, e.g., in physical, biological and social systems~\cite{Pikovsky2001,Osipov2007}. It emerges spontaneously, being enabled by several coupling mechanisms and in the absence of an external driver, differently from entrainment. While it has been well studied in the classical domain~\cite{Arenas2008}, it has recently become a focus of research in the quantum regime~\cite{Galve2017}, where both entrainment~\cite{Goychuk2006,Zhirov2008,Zhirov2009,Lee2013,Walter2014,Sonar2018,Roulet2018a,Sonar2018,Parra2020} and spontaneous synchronization~\cite{Orth2010,Giorgi2012,Ludwig2013,Mari2013,Manzano2013,Hush2015,Xu2014,Cabot2018,Cabot2019,Bellomo2017,Roulet2018,Karpat2019,Eshaki2019,Karpat2020} have been explored in a variety of systems including spins, harmonic and non-linear oscillators, modeling platforms ranging from optomechanical systems to trapped ions and superconducting qubits. The presence of quantum correlations as a signature of synchronization, as well as the origin of these dynamical features has been discussed \cite{Giorgi2012,Manzano2013,Hush2015,Lee2014,Sonar2018}. Signatures of synchronization have also been recently reported in experiments~\cite{Laskar2020,Koppenhfer2019}.

Dissipation is a key enabling mechanism for spontaneous synchronization: indeed  diffusive couplings (cross-damping terms) have been considered in classical systems \cite{Pikovsky2001}, while dissipation, either global  or local, is known to induce synchronization among quantum oscillators and spins, either in the steady state or in the transient relaxation dynamics~\cite{Giorgi2019}. Even if different forms of dissipation and decoherence have been considered, a common feature of these works is the assumption of Markovian evolution of the open quantum systems. Dissipation of quantum systems is mostly described by neglecting memory effects due to its technical simplicity, but non-Markovianity is actually the rule rather than the exception in many realistic settings and there have been several advances in the last decade on the theoretical framework encompassing memory effects. Indeed, quantum non-Markovianity is a multi-faceted phenomenon whose  quantification via various techniques has been vastly explored both theoretically~\cite{Rivas2014,Breuer2016,Li2018,Li2020} and experimentally~\cite{Liu2011,Fanchini2014,Haseli2014}. As memory effects might enable the open system to recover a certain part of the information lost in the environment, they are also known to be relevant in the context of quantum metrology~\cite{PhysRevLett.109.233601}, quantum information processing~\cite{PhysRevLett.108.160402,PhysRevA.99.012319} and thermodynamics~\cite{Bylicka2016,Pezzutto2019}.

Given the enabling role of dissipation for spontaneous synchronization, it is of fundamental and practical interest to establish the effect of memory and non-Markovianity. Our main goal in this work, is to understand the relationship between the degree of non-Markovianity in the open system dynamics and the onset of spontaneous synchronization considering different approaches. We consider a pair of coupled qubits, in a non-symmetric dissipation configuration in which only one is in direct contact with the environment. This configuration has been recently shown to allow for probing of the features of an out-of-equilibrium qubit through measurement of the probe \cite{Giorgi2016} and can be realized, for instance, in atomic platforms \cite{Cabot2019}.  We address the effect of the environment through both a Lindblad master equation and a collision model. The former allows to assess the relation between the local non-Markovianity of one qubit and its ability to synchronize with the other one. The latter allows to go beyond Markovian assumptions for the whole qubit pair open system dynamics. 

One could expect that information backflow on one qubit, being a manifestation of its interaction with the rest of the system and environment, would favour the emergence of synchronization  -- which is normally enhanced by increasing the coupling (see for instance in Refs. \cite{Giorgi2012,Manzano2013}, where Arnold tongue-like phase diagrams \cite{Pikovsky2001} were found). We report on the failure of this intuition.  Using a Lindblad type master equation we show that there exists a trade-off between non-Markovianity of the probe qubit and the emergence of synchronization between the qubit pair. In particular, for a given intraqubit coupling, memory effects tend to be significantly larger where synchronization is absent, as can also be assessed analytically. In accordance, we show that the time required for the establishment of synchronization is related with the inverse of the strength of non-Markovianity. Then, using a collision model, we extend our analysis to a scenario in which the interaction with the environment also gives rise to non-Markovianity through the backflow of information from the environment to the open system, in addition to the intraqubit coupling. Our simulations demonstrate that the trade-off relation between synchronization and non-Markovianity is indeed robust in this more general case as well.

This manuscript is organized as follows. In Sec.~\ref{sec2}, we first introduce the physical setting and the master equation describing the open system dynamics. Then, presenting the figures of merit that we use to quantify the degree of non-Markovianity and synchronization, we report our main results on the trade-off relation between these two concepts. Sec.~\ref{sec3} includes a more general collisional model approach to the same problem, where backflow of information from the environment to the open system is also considered. We conclude in Sec.~\ref{sec4}.

\section{Master Equation and Figures of Merit}\label{sec2}

Let us consider a qubit $s_1$ directly interacting with a second qubit $s_2$, which is  immersed in a boson thermal environment. The total Hamiltonian is $H=H_S+H_B+H_I$, where 
\begin{equation}
H_S=\frac{\omega_1}{2}\sigma_{s_1}^z+\frac{\omega_2}{2}\sigma_{s_2}^z+\lambda (\sigma_{s_1}^+ \sigma_{s_2}^-+\sigma_{s_1}^- \sigma_{s_2}^+)
\end{equation}
describes the free evolution of the two qubits and their direct interaction, $H_B=\sum_k \Omega_k a_k^\dag a_k$ is the bath Hamiltonian, and $H_I=\sum_k g_k (a_k^\dag+ a_k )\sigma_{s_2}^x$ is the interaction between the second qubit and the environment.
The dynamics of the density matrix of the system alone can be analytically calculated, at least in the limit of weak system-bath interaction, by deriving the corresponding Born-Markov master equation~\cite{BreuerPetruccione}. Assuming that the qubit-qubit coupling $\lambda$ is either larger than the system-bath interaction strength or smaller than the absolute value of the detuning $|\Delta|=|\omega_1-\omega_2|$, the open-system dynamics can be described employing a full secular approximation~\cite{Cattaneo2019}. The first step to write such an equation is the diagonalization of $H_S$, that can be written as $ H_S=E_1 (\eta_1^\dag\eta_1-1/2)+E_2 (\eta_2^\dag\eta_2-1/2)$
where $E_1=(\omega_0-R)/2$ and $E_2=(\omega_0+R)/2$, with $\omega_0=\omega_1+\omega_2$, $R=\sqrt{(\omega_1-\omega_2)^2+4\lambda^2}$, and where $\eta_i$ ($\eta_i^\dag$) are fermionic annihilation (creation) operators whose definition in terms of the qubit states is given in App.~\ref{appa}. The corresponding zero-temperature master equation reads
\begin{equation}
     \frac{d\rho(t)}{dt}=-i[H,\rho]+ \Gamma_1\,\sin^2\theta{\cal L}(\eta_1)+ \Gamma_2\,\cos^2\theta{\cal L}(\eta_2),
    \label{me}
\end{equation}
where
\begin{equation}
    \theta=\frac{1}{2}\arctan\frac{2\lambda}{\omega_1-\omega_2},
\end{equation}
${\cal L}(X)=X\rho X^\dag-\{X^\dag X,\rho \}/2$, while $\Gamma_1$ and $\Gamma_2$ are given by the spectral density of the bath, calculated respectively at energies $E_1$ and $E_2$. For the sake of simplicity in the analytical discussion, we will assume a flat spectral density leading to $\Gamma_1=\Gamma_2\equiv \Gamma$. 

In order to establish a quantitative relationship between degree of non-Markovianity and the emergence of spontaneous quantum synchronization, we will make use of the well-known trace distance measure~\cite{Breuer2009} to asses non-Markovianity. In this approach, if the trace distance between two arbitrary initial states of the open system decreases monotonically during the dynamics, we have a memoryless Markovian process. However, if the trace distance undergoes a temporarily increase in certain time intervals throughout the evolution, then there exits a backflow of information from environment to system that represents a signature of non-Markovian memory effects. The trace distance between the two density matrices $\rho_1$ and $\rho_2$ is given by $D(\rho_1, \rho_2)=\frac{1}{2}\textrm{Tr} \left[(\rho_1\!-\!\rho_2)^{\dagger} (\rho_1\!-\!\rho_2)\right]^{1/2}$. Then, the degree of non-Markovianity can be quantified via
\begin{equation}\label{eq:N}
{\cal{N}}_{\max}= \max_{\rho_1(0),\rho_2(0)} \int_{\dot{D}>0}\frac{dD}{dt}dt,
\end{equation}
where optimization is performed over all possible pairs of initial spin states. At this point, we should note that we will also consider an inequivalent non-Markovianity measure in App.~\ref{appc} to demonstrate the generality of our results. 

Spontaneous quantum synchronization between a pair of quantum systems can be said to emerge through the establishment of coherent oscillations in the expectation values of their local observables. While it is generally possible to observe this behavior by just looking at the dynamics of these expectation values, one needs to adopt a measure to quantify the degree of synchronization to be able to make a definite discussion. To this end, we will adopt the well-known Pearson correlation coefficient $C_{12}$, which is a standard tool in statistics for identifying correlations between two data sets, as our figure of merit for the detection of synchronous behavior, as it has been done in the majority of the literature on quantum synchronization~\cite{Galve2017}. Given two discrete variables $x$ and $y$, linear correlation between them can be measured by the Pearson coefficient, which is given as 
\begin{equation}
C_{xy}=
\frac{\sum_t(x_t-\bar{x})(y_t-\bar{y})}{\sqrt{\sum_t(x_t-\bar{x})^2}\sqrt{\sum_t(y_t-\bar{y})^2}}.
\label{eq:pc}
\end{equation} 
Here, $\bar{x}$ and $\bar{y}$ denote the averages of $x$ and $y$ over the data set $t$. $C_{12}$ is a bounded function whose range lies in $[-1, 1]$ and the end points of this range corresponds to two extremes in the synchronization behaviour. In particular, $C_{12}=-1$ and $C_{12}=1$ indicate that two variables under consideration are completely negatively and positively correlated, respectively. To elaborate more on these extremes, fully negative correlation between the variables means that while one of them is increasing the other one is decreasing simultaneously, and fully positive correlation signals that both of them behave in the same way, i.e., they are increasing or decreasing together. 

Translating this to the language of the synchronization phenomenon, as mentioned above, we choose the two variables that go into the calculation of the Pearson coefficient as the expectation values of local observables $\langle \sigma^x_{s_1} \rangle$ and $\langle \sigma^x_{s_2} \rangle$. Consequently, based on the definition of the Pearson coefficient above, completely negative and completely positive correlations imply fully anti-synchronized and fully synchronized behaviors between the local expectation values, respectively. We also note that, despite we focus on a particular observable for concreteness, the emergence of spontaneous synchronization in our study is robust in the sense that it is indeed independent of the choice of specific observables.  In addition, we calculate the expectation values of the local observables $\langle \sigma^x_{s_1} \rangle$ and $\langle \sigma^x_{s_2} \rangle$ in our simulations as discrete samples for the considered time interval. Consequently, as we sample $C_{xy}$ given in Eq. \eqref{eq:pc} over a sliding data window along the total evolution time, we can obtain a time-dependent Pearson coefficient to probe how the oscillations become phase-locked over time. Lastly, to get a smooth behavior in the Pearson coefficient evolution, we allow the adjacent data windows to partially overlap for a certain interval.

\begin{figure}[t]
\centering
\includegraphics[width=0.48\textwidth]{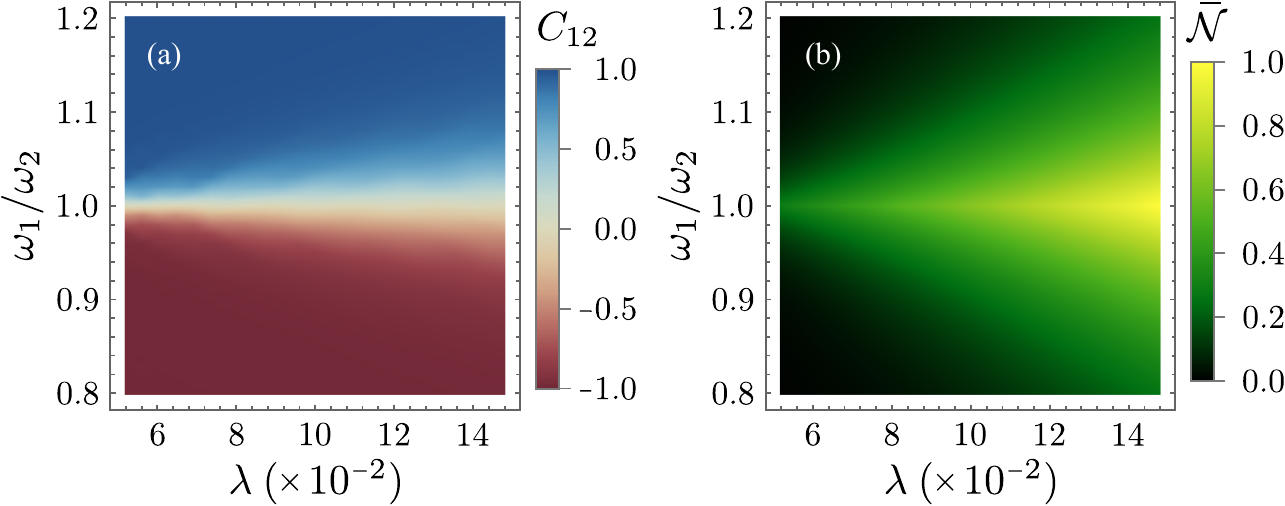}
\caption{(a) Synchronization and (b) normalized non-Markovianity diagrams in terms of the ratio of the self energies of $s_1$ and $s_2$, and the strength of the intra-qubit interactions. In both cases, the system-bath coupling is $\Gamma=0.01$, and the region displayed ensures the validity of the full-secular approximation in Eq. (\ref{me}). In the case of synchronization, the Pearson factor is calculated at time $t\sim 500\; \omega_1$.}
\label{fig1}
\end{figure}

We start by calculating the non-Markovianity of $s_1$ due to the direct coupling to $s_2$ and to the indirect coupling to the environment. To this end, we will prepare both the bath and $s_2$ in their respective ground states. As established in Ref.~\cite{Wismannn2012} in the case of a single qubit, the optimal pair of initial states is represented by a couple of pure, orthogonal states. While in principle one should perform a numerical maximization over all possible pairs of initial states for any value of the system parameters, we have verified that with respect to some given states this only weakly affects the numerical value of the indicator (\ref{eq:N}), but not the landscape of the non-Markovianity dependency of the parameters themselves. So, for the sake of clarity, let us choose the two density matrices $\rho_\pm=|\psi_\pm\rangle\langle \psi_\pm|$ with $|\psi_\pm\rangle=(|0\rangle \pm |1\rangle)/\sqrt{2}$ as the pair for $s_1$. From this point on, we will utilize the symbol $\mathcal{N}$ to label the value of the non-Markovianity measure calculated for the above considered pair of states. In Fig.~\ref{fig1} (b), we show the behavior of $\bar{\mathcal{N}}$ (which is the normalized non-Markovianity obtained by dividing all data points by the maximum value of $\mathcal{N}$) as a function of $\omega_1$ and $\lambda$ (hereafter we will fix $\omega_2=1$ and use it as an energy scale). In order to deeply understand the behavior of Fig.~\ref{fig1} (b), let us try to estimate analytically the value of $\mathcal{N}$. The two initial states for $s_1$ evolve as (see App.~\ref{appb} for details) 
\begin{eqnarray}\label{rhot}
    \rho_\pm^{(1)}(t)&=&\left(\begin{array}{cc}
       p(t)/2  & \pm q(t)/2 \\
        \pm  q^*(t)/2 & 1-p(t)/2
    \end{array}\right)
\end{eqnarray}
with $q(t)=\cos^2\theta e^{-i\frac{(R+\omega_0)}{2}t-\frac{\tilde{\Gamma}_2}{2}t}+\sin^2\theta e^{-i\frac{(R-\omega_0)}{2}t-\frac{\tilde{\Gamma}_1}{2}t}$, where we defined $\tilde{\Gamma}_1=\sin^2\theta\Gamma_1$ and $\tilde{\Gamma}_2=\cos^2\theta\Gamma_2$. The definition of the parameter $p(t)$ is given App.~\ref{appb}. For these two density matrices, the trace distance is given by $D(\rho_+(t),\rho_-(t))=|q(t)|=(\cos^4\theta e^{-\tilde{\Gamma}_1t}+\sin^4\theta e^{-\tilde{\Gamma}_2 t}+2\sin^2\theta\cos^2\theta e^{-\frac{\tilde{\Gamma}_1+\tilde{\Gamma}_2}{2}t} \cos R t )^{1/2}$. While it tends to decrease in time, it can experience partial growths due to  the last term within the square root. Now, $\mathcal{N}$ is given by the sum of $|q(t)|$ calculated over the relative maxima (which can occur at times $t=2 k \pi/R$) minus the sum calculated over the relative minima (at times $t=(2 k-1) \pi/R$).
Increasing the imbalance between the two first terms within the square root, which can be achieved by either increasing the detuning or decreasing the coupling, there are two effects: (i) the difference between $| q(t=2 k \pi/R)|$ and  $|q(t=2 (k-1) \pi/R)|$ is reduced; at the same time (ii) the overall envelop of $|q(t)|$ approaches the zero value faster, so that the number of time intervals that enters the above mentioned sum is also reduced. Both effects cause a fall in $\mathcal{N}$, which explains the behavior of the non-Markovianity plotted in Fig.~\ref{fig1} (b).

Let us now move to the analysis of transient quantum synchronization. As already detailed in the literature~\cite{Galve2017,Giorgi2012,Manzano2013} its emergence is due to the presence of multiple dissipative time scales. If one of these modes is much slower than the others, there is a time window in which the dynamics of all subparties shows a monochromatic oscillation locked at the frequency of that slow mode. Let us remark that the emergence of synchronization is strictly dependent on the spectral properties of the Liouvillian superoperator governing the dynamics, while it is completely independent on the initial state. In our master equation in Eq.~(\ref{me}), such a time-scale separation is expected unless the condition $\cos^2\theta\,\Gamma_1=\sin^2\theta\, \Gamma_2$ is satisfied. Then, under the assumption of a flat density of states ($\Gamma_1=\Gamma_2\equiv\Gamma$), synchronization is expected to be absent along the line $\omega_1=\omega_2$. In Fig.~\ref{fig1} (a), we plot the Pearson coefficient $C_{12}$ calculated at a time long enough  to have all the eigenmodes but the last one  decayed  out (but also shorter than the thermalization time, when also the long-lasting eigenmode would have died) and to observe the emergence of a monochromatic oscillation. The line $\omega_1=\omega_2$ separates a synchronized region from an anti-synchronized one and a different spectral density would lead to a transition line out of perfect qubits resonance for increasing coupling (see Ref.~\cite{Giorgi2016}). Furthermore, the ratio between the two decay rates can be used to estimate the synchronization time. Moving apart from the case $\omega_1=\omega_2$, the separation between the two time scales is a monotonically increasing function of the detuning for any $\lambda$. The difference between the two rates,  $\tilde{\Gamma}_2-\tilde{\Gamma}_1$, for fixed detuning is proportional to $\cos2\theta$, which is monotonically decreasing with $\lambda$. Then, the synchronization time results to be a decreasing function of the detuning and an increasing function of the coupling, being hindered when the first qubit dynamic is non-Markovian. Therefore, the  information backflow on the first qubit (for a Markovian global dissipation of the pair of qubits), does not provide a coupling mechanism beneficial -being actually detrimental- for the emergence of synchronization. 

While the previous analytical arguments strictly apply to the case of a flat spectral density in Eq. (\ref{me}), the correlation between non-Markovianity and synchronization can be traced back beyond such a toy model. Let us consider the more realistic scenario of a system subject to a hybrid noise coming from more than one environment, as discussed for instance in Ref. \cite{Smirnov_2018} where high-frequency and low-frequency noise components coexist, which is typical of platforms employing superconducting qubits \cite{PhysRevLett.100.197001,PhysRevB.83.180502}. Thus, we will now assume the simultaneous presence of two environments, both directly affecting the qubit $s_2$. In order to discuss a realistic situation, we suppose that one of the two baths is Ohmic, that is, it is described by a spectral density $J_{\rm HF}(\omega)=\Gamma_0 \omega e^{-\omega/\omega_c}$, where $\omega_c$ is a (very high) cutoff frequency, which will be neglected in the following treatment, while the second bath is a low-frequency one~\cite{Smirnov_2018}. According to Ref.~\cite{PhysRevLett.100.197001}, in the case of absence of direct tunneling ($\epsilon=0$ in the language employed there), a low-frequency noise causes dephasing in the basis of the system Hamiltonian. Then, assuming statistical independence between the two baths, the master equation given in Eq.~(\ref{me}) is modified as follows:
\begin{eqnarray}
     \frac{d\rho(t)}{dt}&=&-i[H,\rho]+ \Gamma_1\,\sin^2\theta{\cal L}(\eta_1)+ \Gamma_2\,\cos^2\theta{\cal L}(\eta_2) \nonumber  \\
     &+&\Gamma_{\rm LF}(\sigma^z_{s_2}\rho \sigma^z_{s_2}-\rho),
    \label{me1}
\end{eqnarray}
where the couplings to the ohmic bath now are $\Gamma_i=J_{\rm HF}(E_i)$, and  we have introduced a local dissipator for the low-frequency noise, as its amplitude is normally stronger than the one of the high-frequency bath ($\Gamma_{\rm LF}\gg  \Gamma_0$) \cite{Boixo2016}.

\begin{figure}[t]
\begin{center}
\includegraphics[width=0.47\textwidth]{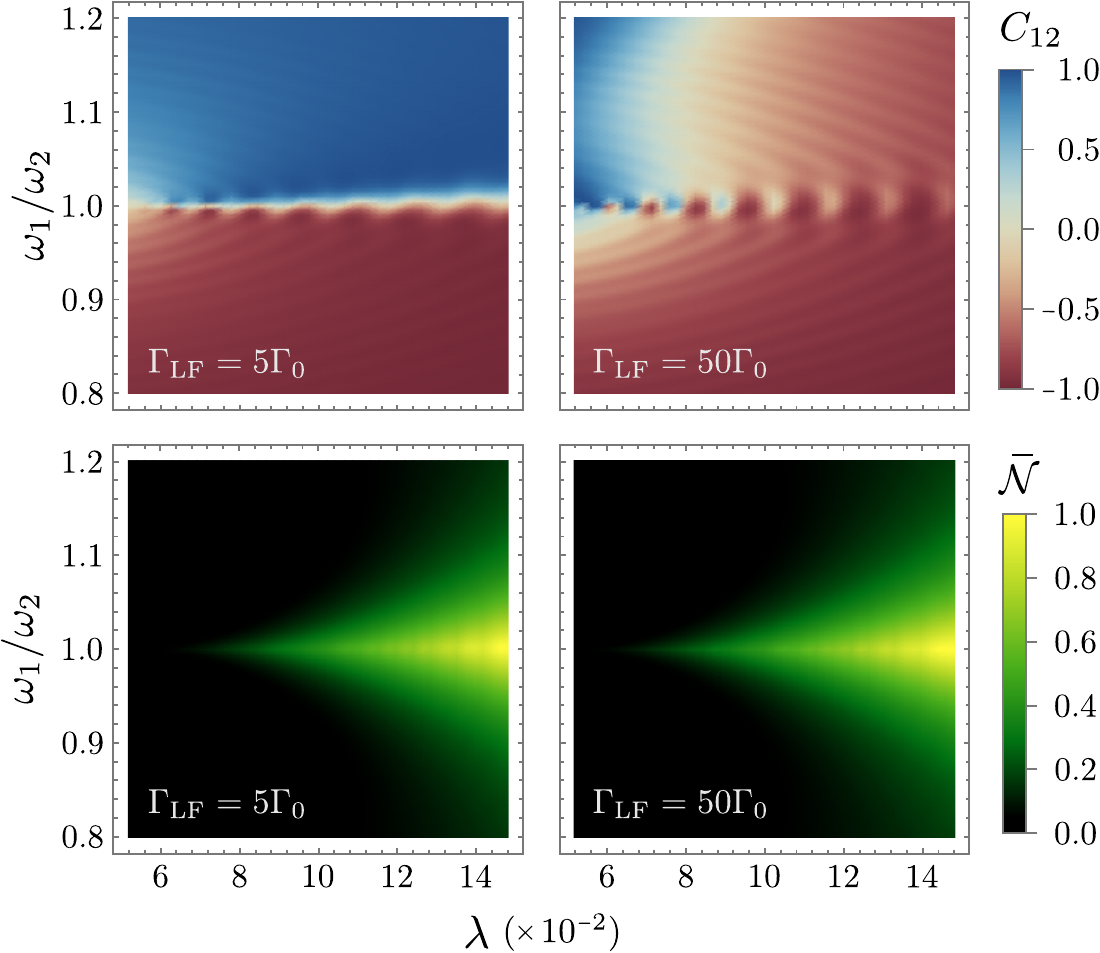}
\end{center}
\caption{Left column: Pearson coefficient (upper row) and normalized non-Markovianity measure (lower row) as a function of the ratio $\omega_1/\omega_2$ and of the coherent coupling in the presence of the hybrid noise described by the master equation in Eq.~(\ref{me1}). The coupling to the bath causing low-frequency noise is $5$ times bigger than the one to the ohmic environment ($\Gamma_{\rm LF}= 0.05 \omega_1$ and $\Gamma_{0}= 0.01\; \omega_1$). Right column: same analysis with the parameter $\Gamma_{0}= 0.001\; \omega_1$. In both cases, the Pearson factor is calculated at time $t\sim 500\; \omega_1$.}
\label{fig2}
\end{figure}

In Fig.~\ref{fig2} we show the phase diagrams of both $C_{12}$ and $\bar{\mathcal{N}}$ as a function of $\lambda$ and of $\omega_1/\omega_2$. As we can see, provided that there is some  amount of dissipation (due to the high-frequency bath), we observe that the general picture is almost indistinguishable with respect to the one showed in Fig.~\ref{fig1} of the main text, which means that the strong correspondence between $C_{12}$ and $\mathcal{N}$ is kept unchanged. An extreme scenario where  such a correspondence is broken takes place considering the limit of very small or even vanishing  coupling to the high-frequency bath. Indeed, in this extreme case, the non-Markovianity phase diagram keeps its structure (also the numerical value of $\mathcal{N}$ is mostly independent), while the synchronization one gets distorted. A possible explanation of this result can be found in previous analysis about the hindering of synchronization in the presence of pure dephasing~\cite{Giorgi2013,Bellomo2017}. In our case, synchronization does not disappear due to the fact that there is direct coupling between the qubits, which implies $[H_S,\sigma_{s_2}^z]\neq 0$. Thus, even with $\Gamma=0$, there is a finite amount of dissipation in the Hamiltonian basis. In the case of $\Gamma_0=0$, a Liouvillian analysis analogous to the one done in Ref.~\cite{Bellomo2017} shows that the slowest decaying mode is always symmetric with respect to the exchange between $\omega_1$ and $\omega_2$, which already excludes the existence of the synchronization/antisynchronization transition.

\section{Collision Model}\label{sec3}

\begin{figure}[t]
\centering
\includegraphics[width=.47\textwidth]{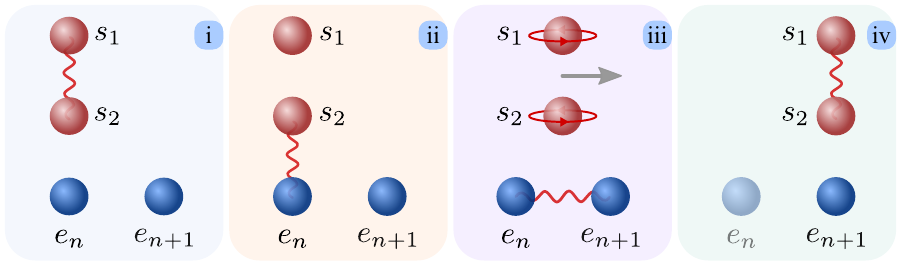}
\caption{Sketch of the model where a single step involves four adjacent collisions. (i) $s_1$ and $s_2$ directly interact with each other. (ii) $s_2$ interacts with the environment particle $e_n$. (iii) As  $s_1$ and $s_2$ freely evolve, a partial SWAP interaction occurs between $e_n$ and $e_{n+1}$. (iv) Finally, $e_n$ is discarded and $s_2$ is now ready to interact with $e_{n+1}$.}
\label{fig3}
\end{figure}

An alternative route to describe the dynamics of open systems is provided by collision models~\cite{scarani2002,ziman2005a,ziman2005b,ziman2010,lorenzo2017b,vacchini2014,ciccarello2017}, which can be used for an exact derivation of the system dynamics, and also provide a highly controllable way of introducing memory effects in the environment~\cite{ciccarello2013,mccloskey2014,strunz2016,filippov2017,cakmak2017,jin2018,campbell2018}. Using this framework, we consider a pair of qubits $s_1$ and $s_2$ in direct interaction with each other. As in the previous set-up, while $s_1$ is isolated from environment, $s_2$ is an open system due to its coupling to an environment, which is made out of identical quantum objects in their ground state. Interactions between the particles occur as successive collisions, i.e., as  pairwise couplings described by unitary operators. In the following, we present the details of a single step in the dynamics as summarized in Fig.~\ref{fig3}.

The scheme begins with the direct interaction of $s_1$ and $s_2$ described by the Hamiltonian $H_{s_1s_2}= \frac{\lambda}{2}(\sigma^{x}_{s_1} \sigma^{x}_{s_2} + \sigma^{y}_{s_1} \sigma^{y}_{s_2})$. Then, $s_2$ interacts with the environment particle $e_n$ through $H_{s_2e_n}= \frac{J}{2}(\sigma^{x}_{s_2} \sigma^{x}_{e_n} + \sigma^{y}_{s_2} \sigma^{y}_{e_n})$. Next, $s_1$ and $s_2$ evolve freely under $H_{s_{1(2)}}=-\frac{\omega_{1(2)}}{2} \sigma^{z}_{s_{1(2)}}$, where $\omega_1$ and $\omega_2$ are the self energies of $s_1$ and $s_2$ respectively. Note that the corresponding evolution operators are given by $U=\exp(-iHt)$ for each Hamiltonian term. At the same time, $e_n$, which has already interacted with $s_2$ previously, interacts with the forthcoming particle $e_{n+1}$ with a partial SWAP operation given by $U_{e_{n+1}e_n}= \cos(\gamma)\mathbb{I}_4+i \sin(\gamma) \text{SWAP}$, where $\mathbb{I}_4$ is the $4\times4$ identity operator and $\gamma$ is the strength of the SWAP operation with $\text{SWAP}=|00\rangle \langle00|+|01\rangle \langle10|+|10\rangle \langle01|+|11\rangle \langle11|$. Lastly, a single cycle is completed tracing out $e_n$ and moving to repeating the above procedure with $e_{n+1}$. In addition to the non-Markovian evolution of $s_1$ caused by the direct interaction between $s_1$ and $s_2$, the presence of intra-environment collisions causes an information backflow to the system qubits, thus providing another source of non-Markovianity. The contribution to the non-Markovianity by the latter mechanism can be controlled by the intra-environment coupling. We set the parameters of our collision model such that $J=1$, $\omega_2=1$, $\delta t_s=\delta t_{s_1s_2}=0.2$ and $\delta t_{s_2e_1}=0.1$. We evaluate $C_{12}$ taking the initial state of $s_1s_2$ as $(|0\rangle+|1\rangle) \otimes (|0\rangle-|1\rangle)/2$. Furthermore, similarly to the master equation description, we calculate $\cal{N}$ supposing that $s_2$  is initially in ground state and the state pair for $s_1$ is $(|0\rangle \pm |1\rangle)/\sqrt{2}$.

\begin{figure}[t]
\centering
\includegraphics[width=0.49\textwidth]{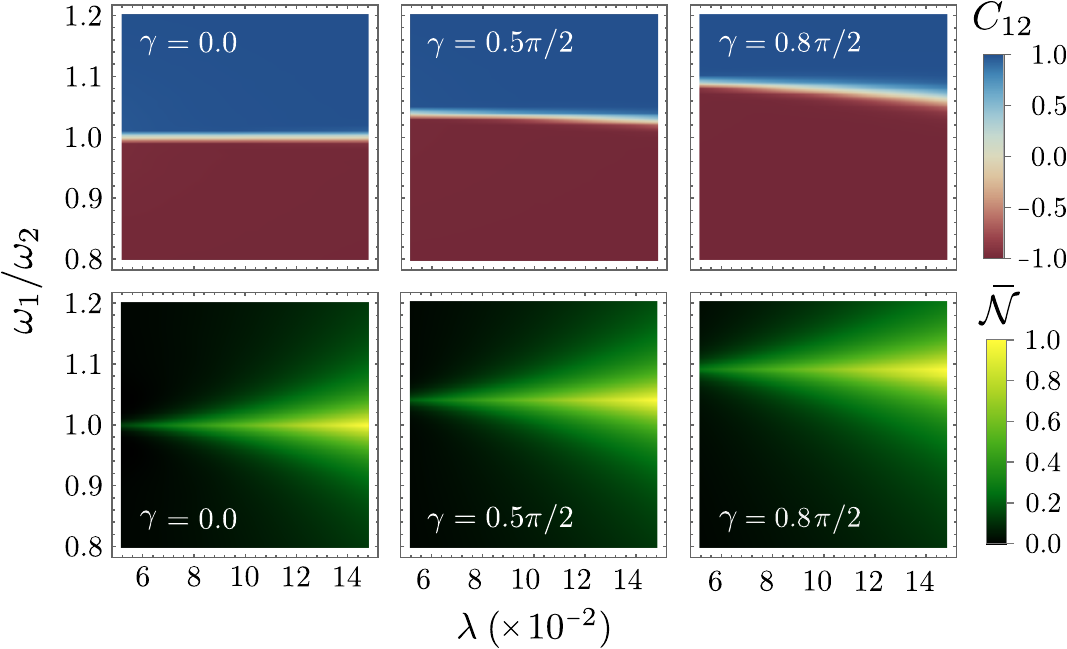}
\caption{While the upper panels display synchronization diagrams for three different intra-environment interaction strengths in terms of the ratio $\omega_1/\omega_2$ of the self energies of $s_1$ and $s_2$ and the strength of the direct interaction between them, the lower panels show the degree of non-Markovianity for the same set of parameters. Here, $C_{12}$ is computed for a sliding data window of 250 collisions, with partial overlaps of 200 collisions, for $N=10000$ iterations.}
\label{fig4}
\end{figure}

\begin{figure}[t]
\includegraphics[width=0.49\textwidth]{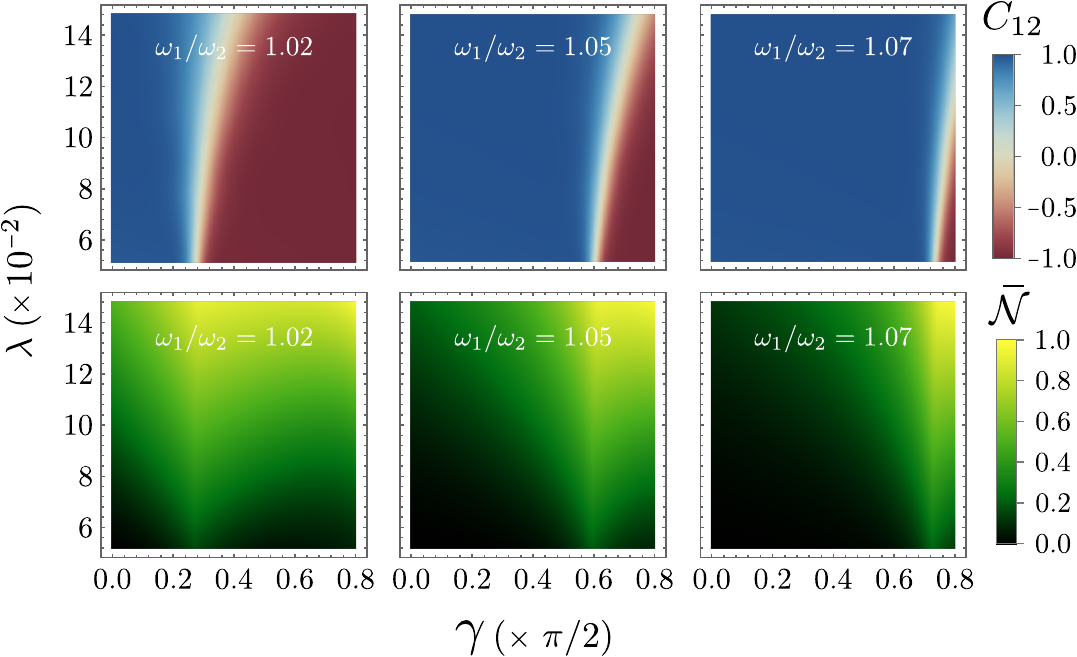}
\caption{As the upper panels show synchronization diagrams for three different $\omega_1/\omega_2$ values for the qubits $s_1$ and $s_2$ in terms of the strengths of the intra-environment and the direct interaction between them, the lower panels display the degree of normalized non-Markovianity for the same set of parameters. The Pearson coefficient $C_{12}$ is computed here for a sliding data window of 250 collisions, with partial overlaps of 200 collisions, for $N=10000$ iterations.}
\label{fig5}
\end{figure}

In Fig.~\ref{fig4}, we present synchronization and first qubit non-Markovianity diagrams as a function of the ratio $\omega_1/\omega_2$ of the self energies of $s_1$ and $s_2$, and the strength of their direct coupling $\lambda$, for three different intra-environment interaction strengths $\gamma$. As can be observed from the upper left panel, when there is no interaction between the environment particles, $\gamma=0$, distinct regions of synchronization and anti-synchronization is sharply separated by the resonance line defined by $\omega_1/\omega_2=1$, in full qualitative agreement with the results obtained considering the Lindblad master equation. Let us stress here that while the master equation has been derived in the weak system-bath coupling, the collision model is built considering a strong coupling, which also shows the robustness of our findings in different regimes. With the second and third upper panels, we demonstrate that as the intra-environment interaction strength $\gamma$ grows stronger, there occurs an upward shift in the sync/anti-sync separation curve. On the other hand, the lower three panels show the corresponding non-Markovianity diagrams for the same three values of $\gamma$. Comparatively analyzing all six diagrams, it is straightforward to see a remarkable trade-off between the degree of memory effects and synchronization, that is, where memory effects become significantly larger, synchronous behavior cannot emerge. Moreover, as will be discussed later, non-Markovianity also has an essential impact on the speed of the establishment of synchronization. Finally, Fig.~\ref{fig5} provides complementary results for our analysis on the interplay between the non-Markovianity of the probe qubit $s_1$ and the emergence of the synchronization between the qubit pair $s_1$ and $s_2$ considering different values of $\omega_1/\omega_2$. The relationship between the asynchronization line defined by the presence of intra-environment collisions and the strength of memory effects becomes evident also in these plots. 

\begin{figure}[t]
\includegraphics[width=.46\textwidth]{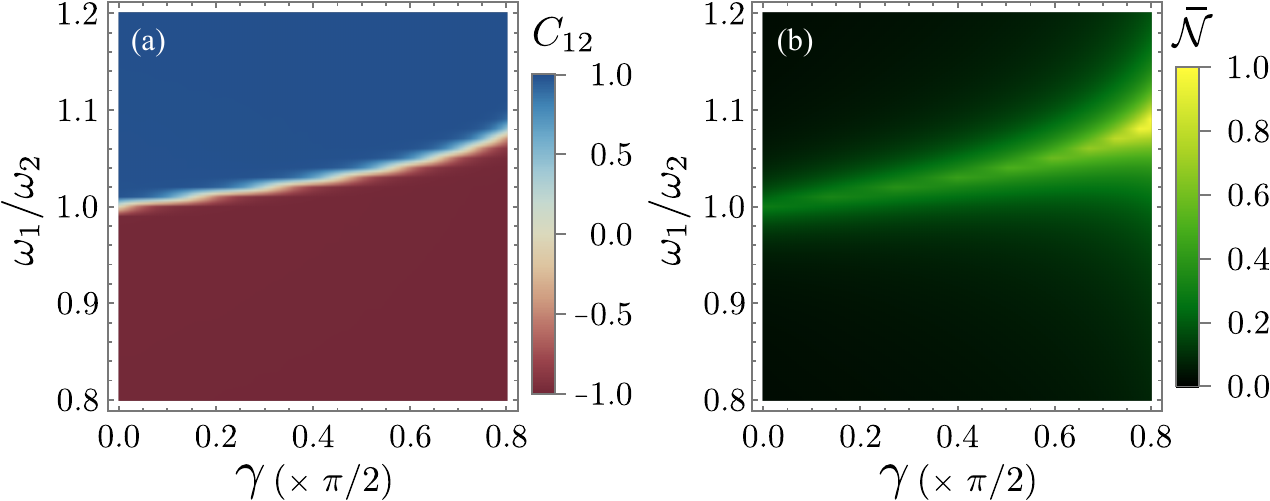}
\caption{(a) Synchronization and (b) normalized non-Markovianity diagrams in terms of the ratio  $\omega_1/\omega_2$ for the self energies of $s_1$ and $s_2$, and the strength of the intra-environment interactions. In both plots, the coupling strength between the particles is $\lambda=0.1$}
\label{fig6}
\end{figure}

Next, we discuss the effects of the intra-environment interaction $\gamma$ and the ratio $\omega_1/\omega_2$ on synchronization and non-Markovianity. In Fig.~\ref{fig6}~(a), we show a synchronization diagram where the coupling strength between $s_1$ and $s_2$ is fixed as $\lambda=0.1$. On the other hand, in Fig.~\ref{fig6}~(b), we display the behavior of $\bar{\mathcal{N}}$ in the dynamics of $s_1$ for the same parameter set. Comparing (a) and (b), we again see a clear relation between the emergence of synchronous behavior between $s_1$ and $s_2$, and the degree of non-Markovianity in the dynamics of $s_1$. Particularly, it can be observed that along the sync/anti-sync separation curve on which synchronization cannot manifest, the degree of memory effects are much larger as compared to the regions where (anti-)synchronization emerges. This conclusion also stands for a different measure of non-Markovianity, for example the entanglement-based measure introduced in~\cite{Rivas2010} as we clearly demonstrate in App.~\ref{appc}.

\begin{figure}[t]
\includegraphics[width=0.48\textwidth]{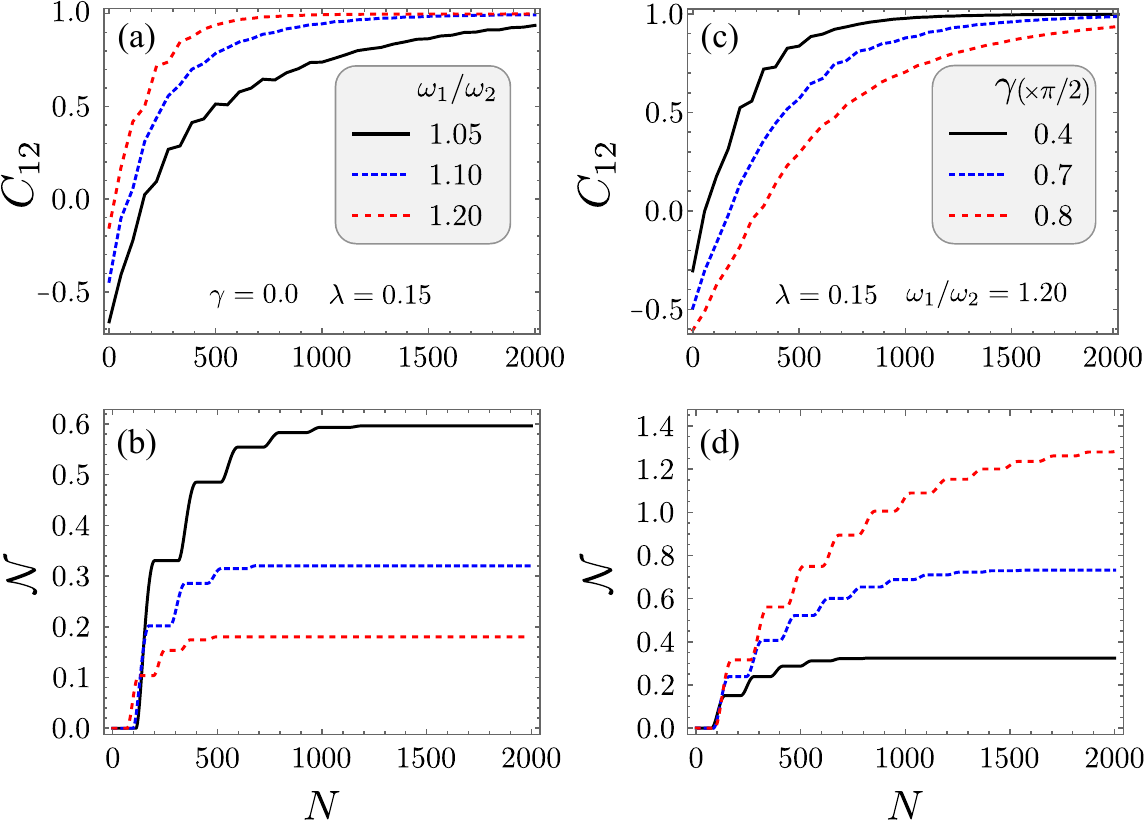}
\caption{Evolution of Pearson coefficient (a) and the degree of non-Markovianity (b) as a function of the number of collisions for three different values of $\omega_1/\omega_2$ between the energies of $s_1$ and $s_2$. Dynamics of Pearson coefficient (c) and non-Markovianity (d) in terms of the number of collisions for three different values of intra-environment interaction strength. Here, we evaluate $C_{12}$ for a sliding data window of 200 collisions with partial overlaps of 150 collisions.}
\label{fig7}
\end{figure}

Lastly, we study the evolution of the Pearson coefficient and non-Markovianity as a function of the number of collisions. In this way, we can better understand how memory effects in the dynamics of $s_1$ affect the speed of the onset of synchronization. In Fig.~\ref{fig7} (a) and (b), we display the dynamics of $C_{12}$ and $\cal{N}$ for three different values of $\omega_1/\omega_2$ when there exist no intra-environment interactions, i.e., $\gamma=0$. In Fig.~\ref{fig7} (c) and (d), we show the evolution of $C_{12}$ and $\cal{N}$ once again but this time assuming fixed detuning ($\omega_1/\omega_2=1.20$) for three different values of intra-environment interaction $\gamma$. Comparing the behavior of $C_{12}$ and $\cal{N}$, it becomes clear that increasing degree of memory effects in the open system dynamics of $s_1$ slows down the emergence of synchronization. 

\section{Conclusion}\label{sec4}

We have provided a comparative analysis of the emergence of spontaneous quantum synchronization and non-Markovian memory effects, defined by the backflow of information from the environment to the open system, and showed that there exists a robust trade-off relation between these two fundamental phenomena. The results have been obtained performing both analytical calculations using a master equation approach (both in a simple single bath with flat spectrum scenario and in a hybrid noise modeling superconductor qubits) and a numerical analysis based on a collision model embedding further memory effects. These two approaches allow to describe the dynamics of the open system in different regimes of the system-bath interaction, which is a further corroboration of the generality  of our results. In particular, our findings point out that as a consequence of the information backflow from the environment to the probe qubit, through its coupling with the open system qubit, appearance of synchronization will be delayed or completely prevented depending on the model parameters. Indeed, the backflow of information on the first qubit does not provide a useful coupling mechanism for synchronization, neither in the weak nor in the strong coupling regime with the environment. The conclusion is robust also considering different non-Markovianity indicators~\cite{Breuer2009,Rivas2010}. Finally, our findings also show that synchronization of the qubit pair in our setting  can be used to probe the degree of non-Markovianity of the open quantum system dynamics.

\section*{Acknowledgements}

G. K. is supported by the BAGEP Award of the Science Academy, the TUBA-GEBIP Award of the Turkish Academy of Sciences, and also by the Technological Research Council of Turkey (TUBITAK) under Grant No. 117F317. \.{I}. Y.\ is supported by M\v{S}MT under Grant No. RVO 14000. B. \c{C}. is supported by the BAGEP Award of the Science Academy, the TUBITAK under Grant No. 117F317, and also by the Research Fund of Bah\c{c}e\c{s}ehir University (BAUBAP) under project no: BAP.2019.02.03. R.Z. and G.L.G. acknowledge financial support from MINECO/AEI/FEDER through projects PID2019-109094GB-C21,  the Severo Ochoa and Mar\'ia de Maeztu Program for Centers and Units of Excellence in R\&D (MDM-2017-0711), CSIC Research Platform PTI-001 and PIE 202050E098, and the QUAREC project funded by CAIB (PRD2018/47).
GLG is funded by the Spanish  Ministerio de Educaci\'on y Formaci\'on Profesional/Ministerio de Universidades   and  co-funded by the University of the Balearic Islands through the Beatriz Galindo program  (BG20/00085). The authors thank the organizers of the QQQ Workshop in Milan, 2020, where the collaboration that lead to this work has initiated.

\appendix

\section{Hamiltonian diagonalization}\label{appa}

Let us consider the system Hamiltonian introduced in Eq. (1) of the main text:
\begin{equation}
H_S=\frac{\omega_1}{2}\sigma_{s_1}^z+\frac{\omega_2}{2}\sigma_{s_2}^z+\lambda (\sigma_{s_1}^+ \sigma_{s_2}^-+\sigma_{s_1}^- \sigma_{s_2}^+)
\end{equation}
Its eigenstates are $|\downarrow\downarrow\rangle,\,|\theta\rangle =\cos\theta |\uparrow\downarrow\rangle+\sin\theta |\downarrow\uparrow\rangle,\,|\theta_\perp\rangle =-\sin\theta |\uparrow\downarrow\rangle+\cos\theta |\downarrow\uparrow\rangle,\,|\uparrow\uparrow\rangle $, with respective energies $-\frac{\omega_1+\omega_2}{2}\equiv -\omega_0/2,\, {\rm sign}(\omega_1-\omega_2)R/2,\,{\rm sign}(\omega_2-\omega_1)R/2,\,\frac{\omega_1+\omega_2}{2}\equiv \omega_0/2$,
where $R=\sqrt{\frac{(\omega_1-\omega_2)^2}{4}+\lambda^2}$ and we have defined
\begin{equation}
  \theta=\frac{1}{2}\arctan\frac{2\lambda}{\omega_1-\omega_2}.
\end{equation}

The Hamiltonian can be put into a quasi-particle form introducing the operators
\begin{eqnarray}
\eta_1^\dag&=& |\theta\rangle\langle\downarrow\downarrow|-|\uparrow\uparrow\rangle\langle\theta_\perp|\\
\eta_2^\dag&=& |\theta_\perp\rangle\langle\downarrow\downarrow|+|\uparrow\uparrow\rangle\langle\theta|
\end{eqnarray} and their respective Hermitian conjugates (it can be verified that such set of operators obeys fermionic anticommutation rules $\{\eta_i,\eta_j\}=0$, $\{\eta^\dag_i,\eta^\dag_j\}=0$, and $\{\eta_i,\eta^\dag_j\}=\delta_{i,j}$).

Using these operators we can rewrite the Hamiltonian as
\begin{equation}
    H=E_1 (\eta_1^\dag\eta_1-1/2)+E_2 (\eta_2^\dag\eta_2-1/2)
\end{equation}
where $E_1=(\omega_0-R)/2$ and $E_2=(\omega_0+R)/2$.
In terms of occupation numbers, the eigenstates are
\begin{eqnarray*}
&&|\downarrow\downarrow\rangle=|00\rangle\\
&&|\theta\rangle =|10\rangle\\
&&|\theta_\perp\rangle =|01\rangle\\
&&|\uparrow\uparrow\rangle=|11\rangle
\end{eqnarray*}

Let us now put the second qubit $s_2$ in contact with a bath through $ H_I=\sum_k g_k (a_k^\dag+ a_k )\sigma_{s_2}^x$.
The operator $\sigma_{s_2}^x$ can be decomposed as 
\begin{equation}
    \sigma_{s_2}^x=\cos\theta (\eta_2^\dag+\eta_2)+\sin \theta(\eta_1^\dag+\eta_1)
\end{equation}
which gives rise to the master equation (2) of the main text, valid under secular approximation and at zero temperature, which we rewrite here:
 \begin{equation}
     \frac{d\rho(t)}{dt}=-i[H,\rho]+ \Gamma_1\,\sin^2\theta{\cal L}(\eta_1)+ \Gamma_2\,\cos^2\theta{\cal L}(\eta_2).\label{mes}
 \end{equation}

\section{Entanglement based non-Markovianity and synchronization}\label{appc}

Quite differently from its classical counterpart, the concept of non-Markovianity is not uniquely defined in the quantum regime. In fact, quantum non-Markovianity is known to be a multifaceted phenomenon which should be studied from many different perspectives~\cite{Breuer2016}. In accordance with this fact, there are now numerous quantifiers in the literature that have been introduced to measure the degree of memory effects in the open system dynamics of quantum systems~\cite{Rivas2014}. It is important to note that almost all of these quantifiers are actually witnesses for the completely positive divisibility of quantum dynamical maps describing the time evolution of open systems. However, many of them have  their own physical motivations which are for instance connected with the dynamics of information flow between the open system and its surrounding environment. Despite the fact that non-Markovianity measures might give inequivalent results under certain conditions (see for example Refs.~\cite{Liu2013,Addis2014,Neto2016}), they are also known to give rise to similar qualitative conclusions in many cases. 

Although we use the first proposed and one of the most established quantifiers to measure the degree of memory effects in open system dynamics in the main text and in other parts of the supplementary material, namely the trace distance measure, here we also present some results on the relation between the emergence of dynamical memory effects and the spontaneous quantum synchronization, considering an alternative measure of non-Markovianity. To this end, we will make use of a correlation based measure of non-Markovian memory effects~\cite{Rivas2010}, which is known to be distinct from the trace distance measure both in terms of its mathematical construction and its physical interpretation. In particular, while the trace distance measure, which is based on the evolution of distinguishability between a pair of states throughout the open system dynamics, can be interpreted to measure the amount of information backflow from the environment to system~\cite{Breuer2009}, the entanglement based measure is directly connected to the information dynamics between an open system and its reservoir through entropic measures~\cite{Fanchini2014}. In other words, even though they might provide similar conclusions in many cases, the two measures are generally inequivalent~\cite{Addis2014}. As will be seen, the results that we obtain using the entanglement based measure confirm the generality of our conclusions on the trade-off relation between the onset of memory effects and the occurrence of spontaneous synchronization. 

\begin{figure}[t]
\begin{center}
\includegraphics[width=0.48\textwidth]{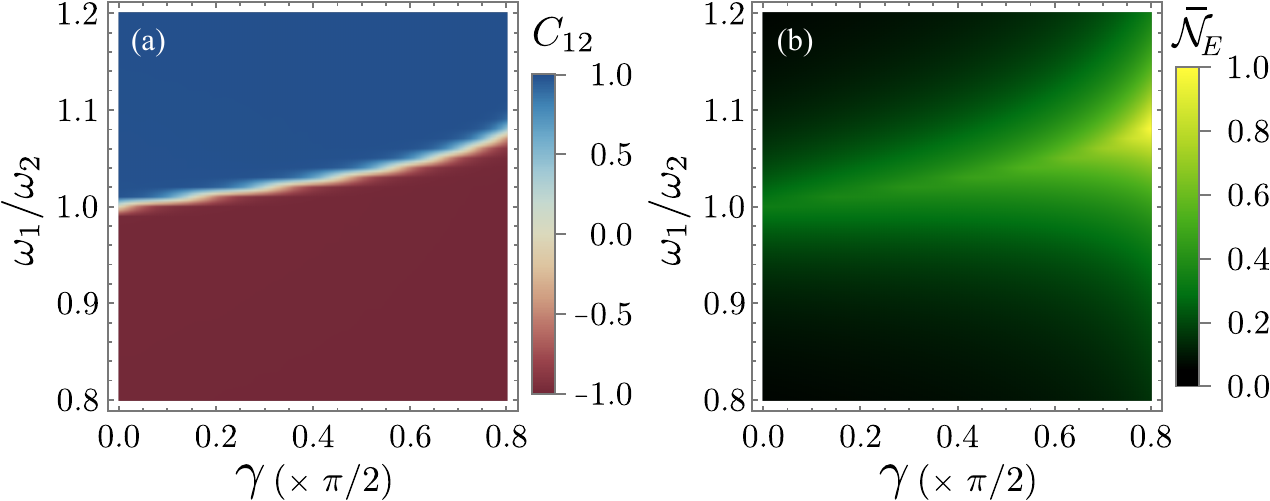}
\end{center}
\caption{(a) Synchronization and (b) normalized entanglement based non-Markovianity diagrams in terms of the ratio $\omega_1/\omega_2$ for $s_1$ and $s_2$, and the strength of the intra-environment interactions. In both plots, the coupling strength between the particles is $\lambda=0.1$}
\label{fig8}
\end{figure}

The alternative measure we consider in this section is constructed upon the entanglement dynamics of a bipartite quantum system that is made up of the principal open system qubit and an additional ancillary qubit which is assumed to be trivially evolving in time without being affected by the environment. In this approach, it is assumed that we introduce an ancillary system $a_1$, having the same dimension as the principal open spin system $s_1$. Then, supposing that the first spin $s_1$ undergoes decoherence due to its direct coupling with the second spin $s_2$, and the ancillary spin $a_1$ evolves trivially, a temporary increase in the entanglement of the bipartite system $a_1s_1$ during the dynamics implies the existence of non-Markovian memory effects. On the other hand, a monotonic decrease in the dynamics of the entanglement means that the open system evolution is Markovian. As a consequence, the degree of memory effects in terms of entanglement can be quantified with the help of the expression~\cite{Rivas2010}
\begin{eqnarray}
\mathcal{N}_E&=&\max_{\rho_{a_1s_1}(0)}\int_{\dot{E}>0}\frac{dE(t)}{dt}dt
\end{eqnarray}
where the optimization must be performed over all possible initial bipartite states $\rho_{a_1s_1}(0)$. As it has been shown that the above quantity is optimized for maximally entangled states~\cite{Neto2016}, we calculate it assuming that the bipartite state $a_1s_1$ is initially in one of the Bell states. Also note that here we choose concurrence to quantify entanglement.

In Fig.~\ref{fig8}, we display a synchronization (a) and an entanglement based non-Markovianity (b) diagram in terms of the intra-environment interaction strength $\gamma$ and the ratio $\omega_1/\omega_2$ of the self energies of the spins. In fact, this figure is identical to the Fig. 4 of our main text, except for the fact that the non-Markovianity diagram here is calculated considering the entanglement based non-Markovianity measure $\mathcal{N}_E$ rather than the trace distance measure $\mathcal{N}$ in the main text. As can be seen from the comparison of these two figures, the results are qualitatively very similar. Therefore, one can clearly see that the demonstrated trade-off relation between the phenomena of spontaneous quantum synchronization and non-Markovianity is not limited to a specific choice of non-Markovianity quantifier, but rather a more general one.

\section{Trace distance based non-Markovianity}\label{appb}

Let us consider the dynamics of the two initial density matrices $\rho_{\pm}=|\psi_{\pm}\rangle\langle \psi_{\pm}| $, with   $|\psi_\pm\rangle= (|\uparrow\rangle\pm|\downarrow\rangle)\otimes|\downarrow\rangle/\sqrt{2}$: in the fermionic  basis we have
\small
\begin{equation}
\rho_\pm(0)=\frac{1}{2}(\cos\theta |10\rangle-\sin\theta  |01\rangle\pm |00\rangle)(\cos\theta \langle 10|-\sin\theta  \langle 01|\pm \langle00|) \nonumber
\end{equation}
\normalsize
Using the master equation (\ref{mes}), we can readily calculate the time evolution of such states, which amounts to
\small
\begin{eqnarray}
\rho_\pm(t)&=&\frac{1}{2}\big\{|00\rangle\langle00|[1+\sin^2\theta(1-e^{-\tilde{\Gamma}_2 t})+\cos^2\theta(1-e^{-\tilde{\Gamma}_1 t})]\nonumber\\
&+&|01\rangle\langle01|\sin^2\theta e^{-\tilde{\Gamma}_2 t }+|10\rangle\langle10|\cos^2\theta e^{-\tilde{\Gamma}_1 t } \nonumber\\
&-& \cos\theta\sin\theta (|01\rangle\langle10|e^{-i R t-\frac{\tilde{\Gamma}_1+\tilde{\Gamma}_2}{2}t}+|10\rangle\langle 01|e^{i R t-\frac{\tilde{\Gamma}_1+\tilde{\Gamma}_2}{2}t}) \nonumber\\
&\mp&\sin\theta e^{-\frac{\tilde{\Gamma}_2}{2}t} (|01\rangle\langle 00|e^{-i\frac{(R+\omega_0)}{2}t}+|00\rangle\langle 01|e^{i\frac{(R+\omega_0)}{2}t})\nonumber\\
&\pm&\cos\theta e^{-\frac{\tilde{\Gamma}_1}{2}t} (|10\rangle\langle 00|e^{i\frac{(R-\omega_0)}{2}t}+|00\rangle\langle 10|e^{-i\frac{(R-\omega_0)}{2}t})
\big\}, \nonumber
\end{eqnarray}
\normalsize
where $\tilde{\Gamma}_1=\Gamma_1\,\sin^2\theta$ and $\tilde{\Gamma}_2=\Gamma_2\,\cos^2\theta$. Performing the trace over the second qubit and moving back to the spin basis,
\small
\begin{eqnarray}
    \rho_\pm^{(1)}&=&\frac{1}{2}\big\{|\downarrow\rangle\langle \downarrow|[1+\sin^2\theta(1-e^{-\tilde{\Gamma}_2 t})+\cos^2\theta(1-e^{-\tilde{\Gamma}_1 t})]\nonumber\\
   &+&\sin^2\theta e^{-\tilde{\Gamma}_2 t}(\sin^2 \theta |\uparrow\rangle\langle \uparrow|+\cos^2\theta |\downarrow\rangle\langle \downarrow|) \nonumber\\
    &+&\cos^2\theta e^{-\tilde{\Gamma}_1 t } (\cos^2 \theta |\uparrow\rangle\langle \uparrow|+\sin^2\theta |\downarrow\rangle\langle \downarrow|)\nonumber\\
   &+& 2\cos^2\theta\sin^2\theta  \cos Rt e^{-\frac{\tilde{\Gamma}_1+\tilde{\Gamma}_2}{2}t}(|\uparrow\rangle\langle\uparrow|-|\downarrow\rangle\langle \downarrow|)\nonumber\\
      &\pm&\sin^2\theta e^{-\frac{\tilde{\Gamma}_2}{2}t} ( |\uparrow\rangle \langle \downarrow|e^{-i\frac{(R+\omega_0)}{2}t}+e^{i\frac{(R+\omega_0)}{2}t}|\downarrow\rangle  \langle \uparrow|)\nonumber\\
    &\pm&\cos^2\theta e^{-\frac{\tilde{\Gamma}_1}{2}t} (\ |\uparrow\rangle )\langle \downarrow|e^{i\frac{(R-\omega_0)}{2}t}+|\downarrow\rangle\langle \uparrow| e^{-i\frac{(R-\omega_0)}{2}t})
  \big\} , \nonumber
\end{eqnarray}
\normalsize
from which we can identify the coefficients $p(t)$ and $q(t)$ given in Eq. (4) of the main text. As the two states only differ in the sign of the nondiagonal elements, their trace distance is simply given by ${\cal D}(\rho_+^{(1)},\rho_-^{(1)})=|q(t)| $.

\bibliography{bibliography}

\end{document}